\documentclass[twocolumn,prl,showpacs,preprintnumbers,amsmath,amssymb,floatfix,superscriptaddress]{revtex4}
\usepackage{graphicx}
\usepackage{amsmath}
\usepackage{amssymb}
\usepackage{color}

\begin{document}

\title{Filler Dependencies of Electrical Conductivity in Nanotube and Nanofiber Composites}

\author{Gianluca Ambrosetti}\email{gianluca.ambrosetti@epfl.ch}\affiliation{LPM, Ecole Polytechnique F\'ed\'erale de Lausanne, Station 17, CH-1015 Lausanne, Switzerland}\affiliation{ICIMSI, University of Applied Sciences of
Southern Switzerland, CH-6928 Manno, Switzerland}
\author{Claudio Grimaldi}\email{claudio.grimaldi@epfl.ch}\affiliation{LPM, Ecole Polytechnique F\'ed\'erale de
Lausanne, Station 17, CH-1015 Lausanne, Switzerland}
\author{Thomas Maeder}\affiliation{LPM, Ecole Polytechnique F\'ed\'erale de
Lausanne, Station 17, CH-1015 Lausanne, Switzerland}
\author{Andrea Danani}\affiliation{ICIMSI, University of Applied Sciences of
Southern Switzerland, CH-6928 Manno, Switzerland}
\author{Peter Ryser}\affiliation{LPM, Ecole Polytechnique F\'ed\'erale de
Lausanne, Station 17, CH-1015 Lausanne, Switzerland}



\begin{abstract}
We report on a model of polymer nanocomposites with fibrous fillers which explicitly considers the microscopic filler features and replicates the composites as random distributions of particles interconnected via electron tunneling. By exploiting the critical path method, we are able to obtain simple formulas, applicable to most nanotube and nanofiber composites, which allow to infer
the overall composite conductivity starting from few parameters like filler volume fraction, size, and aspect-ratio.
The validity of our formulation is assessed by reinterpreting existing experimental results and by extracting the
characteristic tunneling length, which is mostly found within its expected value range. These results can be used practically to tailor the electrical properties of nanocomposites.
\end{abstract}
\maketitle

The inclusion of nanometric conductive fillers such as carbon nanotubes \cite{Bauhofer2009} or nanofibers \cite{Al-Saleh2009} into insulating polymer matrices allows to obtain electrically conductive nanocomposites with unique properties which are widely investigated and have several technological applications ranging from antistatic coatings to printable electronics \cite{Sekitani2009}.
A central issue in this domain is to create composites with an
overall conductivity $\sigma$ which can be controlled through the volume fraction $\phi$, the shape of the
conducting fillers, their dispersion in the polymer matrix, and the local inter-particle electrical connectedness.
Understanding how these local properties affect the composite conductivity is therefore the ultimate goal of
any theoretical investigation of such composites.

A common feature of most random insulator-conductor mixtures is the sharp increase of $\sigma$ once a critical
volume fraction $\phi_c$ of the conductive phase is reached. This transition is generally interpreted in the
framework of percolation theory \cite{Kirkpatrick1973,Stauffer1994,Sahimi2003} and associated with the formation
of a cluster of connected filler particles
that spans the entire sample. The further increase of $\sigma$ for $\phi>\phi_c$ is likewise understood
as the growing of such a cluster.

In the case of composites with a polymeric matrix, the conducting particles are separated from each other
by a thin polymeric layer,
and the conduction between the fillers is originating from quantum mechanical tunneling
processes \cite{Sheng1978,Balberg1987b,Paschen1995,Li2007}. Tunneling has consequently
been implemented in the percolation picture to explain peculiar transport properties near $\phi_c$\cite{Balberg1987b,Vionnet2005,Grimaldi2006,Johner2008}. Nevertheless, although a good understanding of the factors affecting $\phi_c$ exists, a formulation of $\sigma(\phi)$ for $\phi>\phi_c$ with predictive power is still
missing. This especially applies to polymer composites with fillers such as nanotubes
and nanofibers which have diameters which are comparable with the typical distances of tunneling. In these cases, the detailed behavior of tunneling is expected to take an explicit role in $\sigma(\phi)$ and should be taken into account.

In this Letter we address this problem by assuming that the conducting fillers
form a network of globally connected sites via tunneling processes \cite{Ambrosetti2009}.
We show that the calculated conductivity of composites with
different filler aspect-ratios can be re-obtained by the critical path method, allowing for a simple analytical formula for $\sigma(\phi)$, valid for many regimes of interest, and having an explicit dependence on microscopic
quantities such as the geometrical dimensions of the fillers and the tunneling length.
We further show that, by reinterpreting conductivity data of real nanotubes and nanofibers composites, our
formulation leads to consistent estimations of the tunneling length for these systems.

In modeling the conductor-insulator composite morphology, we treat the conducting fillers as identical
impenetrable objects dispersed in a continuous insulating medium, with no interactions between the
conducting and insulating phases.
Furthermore, in order to describe rod-like filler particle shapes, we consider impenetrable spheroids (ellipsoids of revolution) up to the the extreme
prolate ($a/b\gg 1$) limit,
where $a$ and $b$ are the spheroid polar and equatorial semi-axes respectively.
Distributions of spheroids with different
aspect-ratios $a/b$ and volume fractions $\phi$ were generated inside a cubic cell with periodic boundary conditions via random sequential addition \cite{Ambrosetti2008} and successively relaxed through monte-carlo (MC) sweeps \cite{Torquato2001,Ambrosetti2009}. High density configurations were obtained by combining MC sweeps with particle inflation \cite{Torquato2001,Ambrosetti2009}. The absence of global orientational order was verified using the nematic order parameter \cite{Schilling2007}.

\begin{figure*}[t]
    \begin{center}
  \includegraphics[scale=0.85, clip='true']{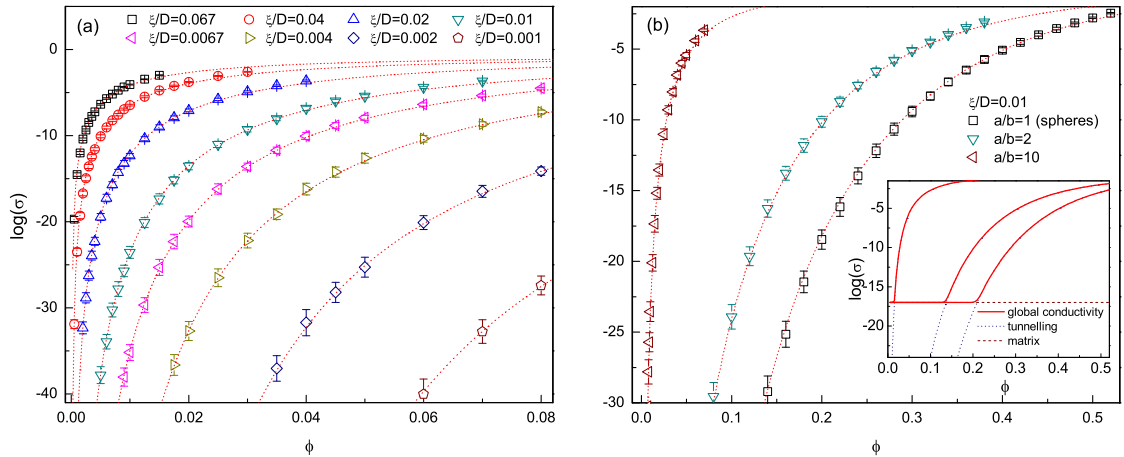}
  \caption{(Color online) (a) Volume fraction $\phi$
  dependence of the tunneling conductivity $\sigma$ for a system of aspect-ratio $a/b=10$ hard prolate spheroids
  with different characteristic tunneling distances $\xi/D$ with $D=2a$. Results from Eq.~\eqref{eq:sigmaSP} (with $\sigma_0=0.179$) are displayed by dotted lines.
  (b) Tunneling conductivity in a system of hard prolate spheroids with different aspect-ratios $a/b$ for the same $\xi/D$. Dotted lines: results from Eq.~\eqref{eq:sigmaSP} with $\sigma_{0}=0.124$ for $a/b=2$, and $\sigma_{0}=0.115$ for $a/b=1$.
  Inset: schematic illustration of the tunneling conductivity crossover for the cases $a/b=1$, $2$, and $10$.}
  \label{Fig1}
  \end{center}
\end{figure*}
In considering the overall conductivity arising in such composites, we assume the spheroids of our
distributions to be perfectly conductive and attribute to each spheroid pair $i,j$ a tunneling
conductance of the form \cite{Balberg1987b}
\begin{equation}
\label{eq:tunneling}
g_{ij}=g_0\exp\!\left(-\frac{2\delta_{ij}}{\xi}\right),
\end{equation}
where $\xi$ is the characteristic tunneling length, which measures the electron wave function decay
within the polymer, and $\delta_{ij}$ is the minimal distance between the two spheroid surfaces. For spheres ($a/b=1$), $\delta_{ij}$ is simply the center-to-center distance minus twice the radius of the two spheres, while for the general case ($a/b\neq 1$), $\delta_{ij}$ depends also on the relative orientation of the spheroids and can be obtained from a numerical procedure described in Ref.[\onlinecite{Ambrosetti2008}].
The specific value of the pre-factor $g_0$ may vary for different composites and, compared to the exponential
term in Eq.~\eqref{eq:tunneling}, has a weaker dependence on $\delta_{ij}$, which we neglect by treating $g_0$ as a constant with unit value.
We also note that in writing Eq.~\eqref{eq:tunneling} we disregarded particle charging effects \cite{Sheng1983,Hu2006} which, at constant room temperature, that is the case considered here, and if the particles are identical and sufficiently large, may be neglected compared to the tunneling decay.

To evaluate the global conductivity of the system, the full set of bond conductances given by equation
Eq.~\eqref{eq:tunneling} is mapped on a resistor network and the overall conductivity is calculated through numerical decimation of the resistor network \cite{Fogelholm1980,Johner2008}. To reduce computational times to manageable limits,
a cutoff distance $\delta_{co}$ is introduced in order to reject negligibly small bond conductances. This was typically equal to $4a$ for low density configurations, while, for spheroids at high densities, it was reduced to $\delta_{co}=2a$ for $a/b=1$, $2$ and $\delta_{co}=a$ for $a/b=10$.

In Fig.~\ref{Fig1}(a) we report the so-obtained conductivity $\sigma$ values (symbols) as
function of the volume fraction $\phi$ of prolate spheroids with aspect-ratio $a/b=10$, and for
different values of $\xi/D$, where $D=2a$. Each symbol is the outcome of $N_R=200$ realizations of a system of $N_P\sim1000$ spheroids. The logarithm average of the results was considered since, due to the exponential dependence of \eqref{eq:tunneling}, the distribution of the computed conductivities was approximately of log-normal form.
The strong reduction of $\sigma$ for decreasing
$\phi$ is a direct consequence of the fact that as $\phi$ is reduced, the inter-particle distances
get larger, leading in turn to a reduction of the local tunneling conductances \eqref{eq:tunneling}.
Furthermore, as shown in Fig.~\ref{Fig1}(b), such reduction depends strongly on
the aspect-ratio of the conducting fillers: as $a/b$ increases, the composite
conductivity drops for much lower values of $\phi$ for fixed $\xi$.
However, in real composites, we have to take into account the
polymer matrix whose intrinsic conductivity $\sigma_m$, which falls typically in the range
$\sigma_m\simeq 10^{-13}\div 10^{-18}$~S/cm, prevents an indefinite drop of $\sigma$.
This is schematically illustrated in the inset of Fig.~\ref{Fig1}(b). Now, if we identify
$\phi_c$ with the volume fraction at which $\sigma\simeq \sigma_m$, then fillers with larger aspect-ratios
entail lower values of $\phi_c$, consistently with what is commonly observed.
This also implies that a model of the composite conductivity based on the tunneling
contribution of Eq.~\eqref{eq:tunneling} alone will be representative only if $\phi$ is sufficiently larger than $\phi_c$ to consider the effect of $\sigma_m$ negligible.

\begin{figure}[t]
    \begin{center}
  \includegraphics[scale=0.9, clip='true']{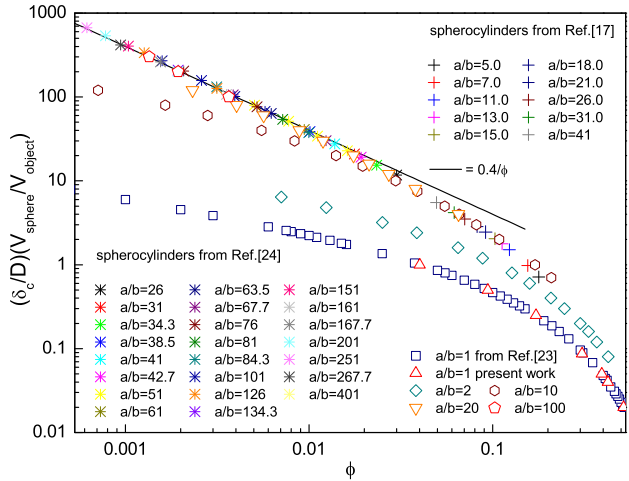}
  \caption{(Color online) Re-scaled critical distances $\delta_c/D$ versus $\phi$ for the prolate spheroids with $a/b=1$, $2$, $10$, $20$, and $100$, and for the impenetrable spherocylinders of Refs.~[\onlinecite{Schilling2007,Berhan2007}]. For $a/b=1$ our results are plotted together with those of Ref.~[\onlinecite{Heyes2006}]. The solid line is the asymptotic behavior for $a/b\gg 1$.}\label{Fig2}
  \end{center}
\end{figure}

We are now going to show that the strong dependence of $\sigma(\phi)$ on $a/b$ and $\xi$ of Fig.~\ref{Fig2}
can be reproduced by the critical path (CP) method \cite{Ambegaokar1971,Shklovskii1984}
applied to our system of impenetrable spheroids. For the tunneling conductances of Eq.~\eqref{eq:tunneling},
this method amounts to keep only the subset of conductances $g_{ij}$ having $\delta_{ij}\leq\delta_c$,
where $\delta_c$, which in turn defines the characteristic conductance $g_c=g_0\exp(-2\delta_c/\xi)$,
is the largest among the $\delta_{ij}$ distances such that the so-defined subnetwork
forms a conducting cluster spanning the sample. Next, by assigning $g_c$ to all the (larger) conductances
of the subnetwork, a CP approximation for $\sigma$ is
\begin{equation}
\label{eq:sigmaSP}
\sigma\simeq\sigma_0\exp\left[-\frac{2\delta_c(\phi,a,b)}{\xi}\right],
\end{equation}
where $\sigma_0$ is a pre-factor proportional to $g_0$.
The significance of Eq.~\eqref{eq:sigmaSP} is that it reduces the conductivity of a distribution of
hard objects electrically connected by tunneling to the computation of the geometrical ``critical"
distance $\delta_c$.
In practice, $\delta_c$ can be obtained by coating each impenetrable spheroid of a distribution with a
penetrable shell of
constant thickness $\delta/2$, and by considering two spheroids as connected if their shells overlap.
$\delta_c$ is then the minimum value of $\delta$ such that, for a given $\phi$, a cluster of connected
spheroids spans the sample. In order to illustrate the efficacy of the CP method, we have calculated
$\delta_c$ for the same parameters of
Fig.~\ref{Fig1} (see below), obtaining an excellent agreement of Eq.~\eqref{eq:sigmaSP} (dotted lines) with
the outcomes of the full numerical decimation of the resistor network.

We shall now illustrate that for sufficiently elongated prolate spheroids, a simple relation exists that allows to estimate $\delta_c$ with good accuracy. In virtue of Eq.~\eqref{eq:sigmaSP} this means that we can formulate explicit relations between $\sigma$ and the shapes and concentration of the conducting fillers.

To extract $\delta_c$ we followed the route outlined in Ref.~[\onlinecite{Ambrosetti2008}]. For given combinations of spheroid aspect-ratios and distance $\delta_c$, we searched the critical (percolation) volume fraction by recording for several $\phi$ the probability of having a cluster spanning the opposite ends of the simulation cell, and identified it with the concentration for which this probability is equal to $1/2$. For every configuration, we used $N_R=40$ for the smallest values of $\delta_c$ up to $N_R=500$ for the largest ones. The particle number varied between $N_P\sim 2000$ ($a/b=1, 2$) and $N_P\sim 8000$ ($a/b=100$). Relative errors on the percolation volume fraction were in the range of a few per thousand.

In Fig.~\ref{Fig2} we report the calculated values of
$\delta_c/D$ as a function of volume fraction $\phi$ for spheres ($a/b=1$, together with the results of
Ref.~[\onlinecite{Heyes2006}]), and for
$a/b=2$, $10$, $20$, and $100$. In the figure, $\delta_c/D$ is multiplied by the ratio $V_{sphere}/V_{object}=(a/b)^2$,
where $V_{sphere}=\pi D^3/6$ is the volume of a sphere with diameter equal to the major axis of the
prolate spheroid and $V_{object}$ is the volume of the spheroid itself.
For comparison, we plot in Fig.~\ref{Fig2} also the results for impenetrable
spherocylinders of Refs.~[\onlinecite{Schilling2007,Berhan2007}]. These are formed by cylinders of radius $R$
and length $L$, capped by hemispheres of radius $R$, so that $a=R+L/2$ and $b=R$, and
$V_{sphere}/V_{object}=(a/b)^{3}/[(3/2)(a/b)-2]\simeq (2/3)(a/b)^{2}$ for $a/b\gg 1$.
As it is apparent, for sufficiently large values of $a/b$ the simple re-scaling transformation collapses both spheroids
and spherocylinders data into a single curve. This holds true as long as the aspect-ratio of the spheroid plus the penetrable
shell $(a+\delta_c/2)/(b+\delta_c/2)$ is larger than about $5$. In addition, for $\phi \lesssim 0.03$
the collapsed data are well approximated by $\delta_c (V_{sphere}/V_{object})/D=0.4/\phi$ (solid line in Fig.~\ref{Fig2}), leading to the following asymptotic formula:
\begin{equation}
\label{eq:deltapro}
\delta_c/D\simeq \frac{\gamma (b/a)^2}{\phi},
\end{equation}
where $\gamma=0.4$ for spheroids and $\gamma=0.6$ for spherocylinders. Equation \eqref{eq:deltapro}
is fully consistent with the scaling law of Ref.~[\onlinecite{Kyrylyuk2008}]
based on the second-virial approximation for semi-penetrable spherocylinders.

We are now in the position to evaluate the tunneling driven
conductivity of random distributions of prolate objects for $\sigma>\sigma_m$.
By substituting Eq.~\eqref{eq:deltapro} into
Eq.~\eqref{eq:sigmaSP} we obtain
\begin{equation}
\label{eq:sigmapro}
\sigma\simeq\sigma_0\exp\!\left[-\frac{2D}{\xi}\frac{\gamma(b/a)^2}{\phi}\right].
\end{equation}
From the previously discussed conditions on the validity of the asymptotic formulas for $\delta_c/D$ it follows
that the above equation will hold when $(b/a)^2\lesssim\phi\lesssim 0.03$. If these conditions are not met, values of
$\delta_c$ can be anyway extrapolated from the data of Figs.~\ref{Fig2}.

\begin{figure}[t]
    \begin{center}
  \includegraphics[scale=1, clip='true']{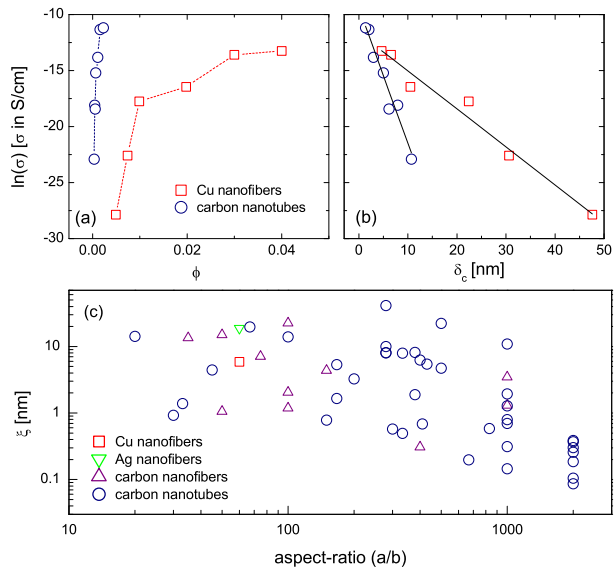}
  \caption{(Color online) (a)
  Natural logarithm of the conductivity $\sigma$ as a function of the volume fraction $\phi$
  for Cu nanofibers-polystyrene\cite{Gelves2006}, and single-wall carbon nanotubes-epoxy\cite{Bryning2005} composites. When, for a given concentration, more then one value of $\sigma$ was
  given, the average of $\ln\sigma$ was considered.
  (b) The same data of (a) re-plotted as function of the corresponding critical distance $\delta_c$.
  Solid lines are fits to Eq.~\eqref{eq:sigmaSPln}.
  (c) Characteristic tunneling distance $\xi$ values for different polymer nanocomposites as extracted by means of Eq.~\eqref{eq:sigmaSPln}}\label{Fig3}
  \end{center}
\end{figure}

Let us now show how the above outlined formalism may be used to re-interpret the experimental data on the conductivity
of different nanocomposites with fibrous fillers that can be found in the literature. In Fig.~\ref{Fig3}(a) we report measured data of $\ln(\sigma)$
versus $\phi$ for polymer composites filled with Cu nanofibers \cite{Gelves2006}, and carbon nanotubes \cite{Bryning2005}. Equation \eqref{eq:sigmaSP} implies that the same
data can be more profitably re-plotted as a function of $\delta_c$, instead of  $\phi$. Indeed, from
\begin{equation}
\label{eq:sigmaSPln}
\ln(\sigma)=-\frac{2}{\xi}\delta_c+\ln(\sigma_{0}),
\end{equation}
a linear behavior with slope $-2/\xi$ is expected, independently of the specific value of $\sigma_0$,
which allows for a direct evaluation of the characteristic tunneling distance $\xi$.
The exact value of $\xi$ is a much investigated topic \cite{Holmlin2001}, and is expected to be between a
fraction of nm and a few nm \cite{Balberg1987b,Shklovskii1984,Holmlin2001,Seager1974,Benoit2002}.
By using the values of $D$ and $a/b$ provided
in Refs. [\onlinecite{Gelves2006,Bryning2005}] and our formula \eqref{eq:deltapro} for $\delta_c$, we
find indeed an approximately linear dependence on $\delta_c$ [Fig.~\ref{Fig3}(b)], from which we
extract $1.65$ nm for the nanotubes, and $5.9$ nm for the nanofibers.

We further applied this procedure to several published data on polymer-based nanotube \cite{Bauhofer2009}
and nanofiber \cite{Al-Saleh2009} composites with fillers
having $a/b$ ranging from $\sim 20$ up to $\sim 2\cdot10^{3}$. 
The results are collected in Fig.~\ref{Fig3}(c), showing that most of the so-obtained values of the tunneling
length $\xi$ are comprised between $\sim 0.1$ nm and $\sim 10$ nm.
This is a striking result considering the number of factors that make a real composite deviate from an
idealized model. Most notably, fillers may have non-uniform size, aspect-ratio, and geometry,
they may be oriented, bent and/or coiled, and interactions with the polymer may lead to agglomeration,
segregation, and sedimentation.
Furthermore, composite processing can alter the properties of the pristine fillers,
e.g. nanotube or nanofiber breaking [which may explain the downward drift of $\xi$ for high aspect-ratios
in Fig.~\ref{Fig3}(c)]. In principle, deviations from ideality can be included in the present formalism
by evaluating their effect on $\delta_c$ \cite{Kyrylyuk2008}. It is however interesting to notice that
all these factors have often competing effects in raising or lowering the composite conductivity, and
Fig.~\ref{Fig3}(c) suggests that in average they compensate each other to some extent, allowing
tunneling conduction to emerge strongly from $\sigma(\phi)$ as a visible characteristic of nanocomposites.

In summary, by extending the critical path method, we have mapped the tunneling driven conductivity $\sigma$ of
dispersions of fibrous conducting fillers into a geometrical percolation problem, which has permitted
us to formulate the filler dependencies of $\sigma$ in terms of a single parameter: the critical distance $\delta_c$.
We have shown that for sufficiently large aspect-ratios of the fillers, $\delta_c$, and so $\sigma$, can
be expressed by a simple formula applicable to real dispersions of nanotubes and nanofibers.
To validate our formulation, we have analyzed published conductivity data for several
nanofibers and nanotube composites and extracted
the corresponding values of the tunneling length $\xi$, which is mostly found within its expected range. The above outlined procedures can be likewise used as guidelines to tailor the electrical properties of real composites, and can be generalized to include different filler shapes, filler size and/or aspect-ratio polydispersity, and interactions with the polymer.

This study was supported by the Swiss Commission for Technological Innovation (CTI) through project GraPoly, (CTI Grant No. 8597.2), a joint collaboration led by TIMCAL Graphite \& Carbon SA, and partly by the Swiss National Science
Foundation (Grant No. 200021-121740). Discussions with E. Grivei and N. Johner were greatly appreciated.

\pagebreak

\end{document}